\documentclass[aps,preprint]{revtex4}
\usepackage{graphicx}
\begin{document}
\title{\Large \bf Stationary Spinning Strings and Symmetries of Classical Spacetimes}
\author{\large Haji Ahmedov and Alikram N. Aliev }
\address{Feza G\"ursey Institute, P. K. 6  \c Cengelk\" oy, 34684 Istanbul, Turkey}
\date{\today}

\begin{abstract}

We explore the symmetries of classical stationary spacetimes in terms of the dynamics of a spinning string  described by a worldsheet supersymmetric action. We show that for stationary configurations of the string,  the action reduces to that for a pseudo-classical spinning point particle in an {\it effective} space, which is  a conformally scaled quotient space of the original spacetime. As an example, we consider the stationary spinning string in the Kerr-Newman spacetime, whose motion is equivalent to that of the  spinning point particle in
the three-dimensional effective  space.  We present the Killing tensor as well as the spin-valued Killing vector of this space. However, the nongeneric supersymmetry corresponding to the Killing-Yano tensor of
the Kerr-Newman spacetime  is lost in the effective space.

\end{abstract}

%\pacs{04.20.Jb, 04.70.Bw, 04.50.+h}

\maketitle

\section{Introduction}

The equilibrium  configurations of a Nambu-Goto string in a general stationary  four-dimensional spacetime  were first studied by Frolov {\it et al} \cite{fszh}. It was shown that for the {\it stationary}  string, when the timelike Killing vector of the spacetime  is supposed to be tangent to the string  worldsheet, the Nambu-Goto action reduces to that  for a classical test particle in an effective three-dimensional space. In other words, the problem of the stationary string motion in the original spacetime amounts to the study of geodesics of the effective space, which is one dimension fewer. Using this remarkable fact for the stationary string motion near a Kerr-Newman black hole, the authors of  \cite{fszh} demonstrated that   the Nambu-Goto equation of motion is completely integrable. This was another manifestation of the Kerr-Newman spacetime-``miracle", earlier discovered by Carter \cite{carter} for the integrability of the ordinary  Hamilton-Jacobi equation in this spacetime. In both cases, the integrability originates from the hidden symmetries of the Kerr-Newman spacetime generated by a second rank Killing tensor \cite{wp}.

In recent studies, it has been shown that these properties  also survive in higher dimensions: The most general metrics for rotating  black holes in all higher dimensions possess hidden symmetries, admitting the Killing  and the Killing-Yano tensors \cite{fk1, fk2}.  It has also been shown that the higher-dimensional black hole spacetimes admit the complete separation of variables in the Nambu-Goto equation for  stationary  string configurations \cite{fk3,fk4}. This fact is related to the existence of the Killing tensor just as in the case of  four-dimensional Kerr-Newman spacetime \cite{fszh}. The hidden symmetries of rotating charged black holes  in five-dimensional minimal gauged supergravity \cite{cclp} and the integrability of geodesics for the corresponding Hamilton-Jacobi and Nambu-Goto equations  have been studied in \cite{dkl, ad, aa1}.

An important manifestation of the hidden symmetries also occurred in separation of the Dirac equation in the spacetime of rotating black holes in four dimensions \cite{ chandra, guven}. It is the Killing-Yano tensor \cite{penrose} which lies at the root of this separability. The authors of work \cite{cartermc} were able to verify this fact by  constructing a linear differential  operator, which depends on the Killing-Yano tensor and commutes with the Dirac operator.  Later on,
the systematic exploration of these aspects of the hidden symmetries, in the light of their relation to {\it supersymmetry}, was undertaken in  \cite{gibbons}. In particular, it was shown that the existence of the second rank Killing-Yano tensor in the Kerr-Newman metric implies a new nontrivial supersymmetry (``hidden" supersymmetry) in the motion of  a pseudo-classical spinning point particle in this metric. The similar analysis of the hidden  supersymmetry in the metrics of higher-dimensional rotating black holes is  given in \cite{aa2}.

The purpose of this paper is to study the symmetries of  stationary spacetimes  in terms of the motion of a stationary spinning string described by a worldsheet supersymmetric action. In Sec.II we briefly recall the main facts about the action of a spinning string in a  curved spacetime. In Sec.III we show that, when both the curved spacetime  and the spinning string  are stationary,
the string action reduces  to that for the pseudo-classical spinning point particle in the effective space, which is conformally related to the quotient space of the original spacetime. This precisely resembles the fact mentioned at the beginning  for the stationary string motion governed by the usual bosonic Nambu-Goto action. In Sec.IV we give  the basic equations describing the symmetries and  conserved quantities
in  supersymmetric mechanics of the pseudo-classical spinning particle in the effective space. In Sec.V, as an illustrative example, we consider the motion of the stationary spinning  string in  the Kerr-Newman spacetime.

\section{The Action for a Spinning String}

As is known \cite{brink, deser, green}, the action for the spinning string is obtained by an extension of the action for a relativistic bosonic string, where in addition to the usual position variable $ x ^{\mu}(\zeta^{\alpha})$ one also introduces a fermion  variable $ \psi ^{\mu}(\zeta^{\alpha})$. Here   $ \zeta^{\alpha} = (\zeta^0,\, \zeta^1)$ are  coordinates on the worldsheet of the string , the coordinate function $ x ^{\mu} $  is a spacetime vector and $ \psi ^{\mu} $  is a two-dimensional  Majorana spinor  on the worldsheet, which also behaves as a spacetime vector. In the following, for convenience of the notation, we omit all spinor  indices. The spacetime index $ \mu $ runs over values $ (0, \, 1,....,\textrm{D}-1) $, where $ \textrm{D}  $ is the dimension of the spacetime. In a general curved spacetime with  metric $ g_{\mu\nu}(x)  $ the action is given by
\begin{eqnarray}
S &= & -\frac{1}{2}\int d^2\zeta \sqrt{-G} \left(
G^{\alpha\beta}\partial_\alpha x^\mu \partial_\beta x^\nu g_{\mu\nu}
-i \overline{\psi}^{A}\rho^{\alpha}D_{\alpha}\psi_{A}\right)\,,
\label{actionSS}
\end{eqnarray}
where $ G $ is the determinant of the worldsheet  metric $ G_{\alpha\beta} $. The description of spin degrees of freedom in the curved spacetime  makes it necessary to introduce  both  a `vielbein'  $ e^{\,A}_\mu(x) $ and a `zweibein' $ E^{\,a}_\alpha(\zeta) $ fields, such that
\begin{eqnarray}
g_{\mu\nu}& = & \eta_{AB} e^{\,A}_\mu \,e^{\,B}_\nu\,,~~~G^{\alpha\beta} = \eta^{ab} E^{\alpha}_{\,a} \,E^{\beta}_{\,b}\,,~~~\psi^\mu= e^\mu_ {\,A} \,\psi^A\,,~~~\rho^{\alpha}= E^{\alpha}_{\,a }\rho^{a}\,\,,
\label{vzw}
\end{eqnarray}
where the indices  $ A $ and $ a $  are the local Lorentz indices, $ \eta_{AB} $  is the Minkowski  metric of  the tangent space to the spacetime manifold, $ \eta_{ab} $ is a two-dimensional Minkowski  metric on the string worldsheet. The quantity  $ \rho^{a} $  defines the usual Dirac matrices in two dimensions
\begin{equation}
    \rho^0=\left(
             \begin{array}{cc}
               0 & -i \\
               i & 0 \\
             \end{array}
           \right)\,,~~~~~~ \ \ \\
\rho^1=\left(
             \begin{array}{cc}
               0 &\, i \\
               i & \,0 \\
             \end{array}
           \right)\,,
\label{diracmtx}
\end{equation}
obeying the relation
\begin{equation}
\{ \rho^a, \rho^b\}= -2 \eta^{ab}\,
\label{anticomrel}
\end{equation}
and $ \overline{\psi}= {\psi}^\dag \rho^0 $. The  covariant derivative is given by
\begin{equation}
D_\alpha \psi_A =\partial_\alpha \psi_A - \partial_\alpha x^\mu
    \omega_{\mu AB} \psi^B\,,
\end{equation}
where $\omega_{\mu AB}$ is the spin connection on the spacetime manifold, while the spin connection on the worldsheet does not contribute to this expression.

The gauge invariance properties of the action (\ref{actionSS}) that is, its  reparametrization symmetries and the local Weyl symmetry allow one to make a particular choice for the metric functions $ G_{\alpha\beta} $. The most convenient choice is achieved in a conformal gauge, in which one has a flat metric  on the worldsheet. In this gauge, the action (\ref{actionSS}) reduces to the form
\begin{equation}
S=-\frac{1}{2}\int d^2\zeta \left(\eta^{a b}
\partial_{a}  x^\mu \partial_{b} x^\nu g_{\mu\nu} - i
\overline{\psi}^{A} \rho^{a} D_{a} \psi_{A} \right)\,.
\label{conactionSS}
\end{equation}
This action possesses  a global worldsheet supersymmetry relating bosonic and fermionic coordinates (see for instance, Ref.\cite{green} for details). Applying to this action the Noether procedure and using the transformations of the worldsheet supersymmetry,  we obtain the two-dimensional energy momentum tensor
\begin{eqnarray}
T_{a b }&= & g_{\mu\nu} \partial_{a} x^\mu \partial_{b} x^\nu -\frac{i}{2}\,\overline{\psi}^{A} \rho_{(a} D_{b)}\psi_{A} - \frac{\eta_{a b}}{2}\left(g_{\mu\nu}\partial^c x^\mu \partial_c x^\nu - \frac{i}{2}\,\overline{\psi}^{A}\rho^{c}
D_{c}\psi_{A}\right)\,
 \label{emt}
\end{eqnarray}
and the supercurrent
\begin{equation}
Q_{a} = \frac{1}{2}\,\rho^{b}\rho_{a} \psi_\mu  D_{b} x^\mu\,\,.
\label{scurrent}
\end{equation}
The round parentheses here and in the following denote symmetrization over the indices enclosed. The equations of motion for $ x ^{\mu} $  and $ \psi ^{\mu} $ derived from the action (\ref{conactionSS}) are the familiar Nambu-Goto and Dirac equations in the curved background. They must be supplemented by the  constraints that the energy momentum tensor (\ref{emt}) and the supercurrent (\ref{scurrent}) vanish on the worldsheet. The consistent  derivation of these constraints  requires the invariance of the action for the spinning string under local supersymmetry as well. This is achieved by adding to the action (\ref{actionSS}) an auxiliary Rarita-Schwinger field $ \chi _{\alpha}(\zeta)$, a fermionic  partner to the zweibein field $ E^a_\alpha(\zeta) $  \cite{brink, green}. Large enough number of the local bosonic and fermionic symmetries of the resulting action enables one to fix the gauge $ E^{\,a}_\alpha = \delta^{\,a}_\alpha\,,~~   \chi_{\alpha}=0 $. With this gauge, we again arrive at the action of the form (\ref{conactionSS}). Furthermore, the variation of the local supersymmetric action with respect to the fields $ E^{\,a}_\alpha $ and $ \chi _{\alpha} $, evaluated in this gauge, results in the constraint equations
\begin{eqnarray}
T_{a b }&= & 0\,,~~~~~~~~~ Q_{a} = 0\,,
\label{constreqs}
\end{eqnarray}
which accompany the equations of motion for the physical fields $ x ^{\mu} $  and $ \psi ^{\mu} $.  In the first case, we  have two independent equations because of the traceless nature of the energy momentum tensor, while in the latter case  there exists one independent equation  due to the identity  $ \rho^{a} Q_{a}=0 $.

\section{Stationary Spinning Strings}

Let us now suppose that the spacetime metric $ g_{\mu\nu} $ in the action (\ref{conactionSS})  admits a timelike Killing vector $ \xi= \xi^{\mu}\, \partial_{\mu} $. That is, the spinning string moves in a stationary curved spacetime $ M $. In this case to proceed further it is useful to employ the formalism of Geroch \cite{geroch} based on a foliation of the spacetime  by its Killing trajectories. Assuming that the timelike Killing vector is  not hypersurface orthogonal, one can consider a set of the Killing trajectories as a quotient space $\cal{M} $ of the  spacetime  $ M $ ($ {\cal{M}}=M/G_1 $, where  $ G_1 $ is a one-dimensional group generated by $ \xi $). Then the metric of the quotient space  $ h_{\mu\nu} $ is related to the spacetime metric  by
\begin{equation}
g_{\mu\nu}=
h_{\mu\nu} +\frac{\xi_\mu\xi_\nu}{\xi^2}\,,
\label{decomp}
\end{equation}
where $ \xi^2= g_{00} $,  and for the corresponding contravariant components we have
\begin{equation}
g^{\mu\nu} =
\begin{pmatrix}
{\xi^{-2}\left(1+h^{ij}\xi_i \xi_j/\xi^2 \right) && - h^{ij}\xi_j/\xi^2 \cr\cr - h^{ij}\xi_i/\xi^2
 && h^{ij}}
\end{pmatrix} .
\label{invmetric}
\end{equation}
We note that the metric $ h_{ij} $ satisfies the completeness relation
\begin{equation}
h^{ik}\,h_{kj} = \delta_{j}^i\,\,, ~~~~ i\,,j= 1,...,\textrm{D}-1\,.
\label{completeness}
\end{equation}
The remarkable feature of this foliation is that  there exists  a uniquely ``mirrored correspondence" between the differential geometries in the original spacetime and in its quotient space. The projection operator
\begin{equation}
h_{\mu}^{\nu}=
\delta_{\mu}^{\nu} -\frac{\xi_\mu \xi^\nu}{\xi^2}\,,
\label{projector}
\end{equation}
provides this correspondence for any stationary tensor field, $  (\mathcal{L}_\xi T=0 $, where $ \mathcal{L}_\xi $ is the Lie derivative along $ \xi $) and its covariant derivative
\begin{eqnarray}
{\hat T}^{\nu_1\cdots \nu_n}_{~~\mu_1\cdots \mu_m}&= & h^{\nu_1}_{~\lambda_1}\cdots
h^{~\rho_m}_{\mu_m}\, T^{\lambda_1\cdots \lambda_n}_{~~\rho_1\cdots \rho_m}\,,~~~
{\hat D}_{\gamma}{\hat T}^{\nu_1\cdots \nu_n}_{~~\mu_1\cdots \mu_m}= h^{\nu_1}_{~\lambda_1}\cdots
h^{~\rho_m}_{\mu_m}\,h^{~\kappa}_{\gamma} D_{\kappa}{\hat T}^{\lambda_1\cdots \lambda_n}_{~~\rho_1\cdots \rho_m},
\label{corresp}
\end{eqnarray}
where the covariant derivative operators  $ {\hat D}_{\mu } $  and $  D_{\mu } $ are associated with metrics  $  h_{\mu \nu} $ and   $  g_{\mu\nu } $, respectively. Further details can be found in \cite{geroch}.

Next, we also suppose that not only the spacetime, but also  the spinning string itself is {\it stationary}.  For this time independent configuration, the embedding of  the string worldsheet into the spacetime $ M $ can be parameterized by the equations
\begin{equation}
x^0=\tau+ f(\sigma)\,,~~~~~ x^i=x^i(\sigma)\,,
\label{embedd}
\end{equation}
where we have used $ \zeta^0 = \tau $  and  $ \zeta^1 = \sigma $. That is,  the basis vector $\partial_\tau x^\mu $  in the tangent plane to the worldsheet is parallel to the timelike  Killing vector $ \xi^{\mu} $. As for the other tangent vector to the worldsheet, $ \eta^{\mu} = \partial_\sigma x^\mu $,  we require it to lie in the tangent plane of the  quotient space $\cal{M} $, so that
\begin{equation}
\xi_\mu \eta^\mu=0\,.
\label{orthonor}
\end{equation}
Using constraints  (\ref{embedd}) and (\ref{orthonor}) along with  (\ref{projector}) one can easily take the projection of the bosonic term in the action (\ref{conactionSS}) onto  the  quotient  space. We have
\begin{eqnarray}
\eta^{a b}
\partial_{a}  x^\mu \partial_{b} x^\nu g_{\mu\nu} & = &  -\xi^2 + h_{\mu\nu} \eta^\mu \eta^\nu\,.
\label{bosterm}
\end{eqnarray}
The spinor term in  the action (\ref{conactionSS})  can be written in the form
\begin{equation}
\overline{\psi}^\mu \rho^{\alpha}  D_{\alpha}\psi_\mu= \xi^\nu\, \overline{\psi}^\mu \rho^0 D_\nu
\psi_\mu + \eta^\nu \, \overline{\psi}^\mu \rho^1 D_\nu \psi_\mu\,.
\label{ferm}
\end{equation}
We require that the spinors obey the conditions
\begin{equation}
\mathcal{L}_\xi \psi_\mu =0\,,~~~~~~ \psi_\mu \xi^\mu
=\xi^2 \Upsilon\,,
\label{spinconstr1}
\end{equation}
where $\Upsilon$ is a constant spinor. With these conditions we have the decomposition
\begin{equation}
\psi_\mu =\chi_\mu +\xi_\mu \Upsilon\,,
\label{decompose}
\end{equation}
where $\chi_\mu $ is a spinor field on the  quotient space, $ \chi_\mu \xi^\mu=0 $. Substituting this decomposition in equation (\ref{ferm})  and taking into account the conditions (\ref{spinconstr1}) along with the mirrored correspondence (\ref{corresp}),  we obtain that
\begin{equation}
\overline{\psi}^\mu \rho^{a}D_{a}\psi_\mu= \eta^\nu \, \overline{\chi}^\mu \rho^1 {\hat D}_\nu
\chi_\mu + 2 \xi^{\nu}D_\nu \xi_\mu \,\overline{\chi}^\mu \rho^0
\Upsilon\,.
\label{ferm1}
\end{equation}
Thus, the action (\ref{conactionSS}) takes the form $ S= - I \Delta \tau $, where
\begin{eqnarray}
I&=&\frac{1}{2}\int d\sigma \left( -\xi^2 +
h_{\mu\nu} \eta^\mu \eta^\nu
- i \eta^\nu \, \overline{\chi}^\mu \rho^1 {\hat D}_\nu
\chi_\mu -2 i \xi^\nu D_\nu\xi_\mu \overline{\chi}^\mu
\rho^0 \Upsilon \right)\,.
\label{redaction}
\end{eqnarray}
Similarly, the constraint equations accompanying this action are obtained by projecting equations (\ref{constreqs}) with (\ref{emt}) and  (\ref{scurrent})  onto the  quotient  space. We have $ \hat{T}_{01}=\hat{T}_{10}\equiv0 $
and
\begin{eqnarray}
\hat{T}_{00} = \frac{1}{2}\left( \xi^2+ h_{\mu\nu} \eta^\mu \eta^\nu \right) &= & 0\,,~~~~~~
\hat{Q}_0 =\frac{1}{2}\left(\xi^2 \Upsilon + \rho^0\rho^1 \chi_\mu \eta ^\mu\right) = 0\,.
\label{emtproj}
\end{eqnarray}
Next, we introduce an ``effective" metric
\begin{equation}
H_{\mu\nu} =-\xi^2 h_{\mu\nu}\,,
\label{eff}
\end{equation}
which is a  conformally adjusted metric of the quotient space and a new variable $ d\sigma^{\prime} = -\xi^2 d\sigma $. Then it is straightforward to show that the action in (\ref{redaction}) can be put in the form
\begin{equation}
I=\frac{1}{2}\int d\sigma^{\prime}\left(
   H_{\mu\nu} {\eta^{\prime}}^\mu {\eta^{\prime}}^\nu
   - i {\eta^{\prime}}^\nu \overline{\chi}^\mu \rho^1 D^\prime_\nu
\chi_\mu \right)\,,
\label{primeaction}
\end{equation}
where $ {\eta^{\prime}}^\mu = \partial_{\sigma^\prime} x^\mu $ and $ D^\prime_\nu $ is the covariant derivative operator with respect to the metric $ H_{\mu\nu} $. In obtaining this expression we have used the relation
\begin{equation}
\eta^\nu \overline{\chi}^\mu \rho^1 \left(D^\prime_\nu -
\hat{D}_\nu\right)\chi_\mu= 2  \xi^\nu D_\nu\xi_\mu
\overline{\chi}^\mu\rho^0 \Upsilon\,,
\label{confrel}
\end{equation}
which is derived by direct evaluating  the actions of the covariant derivative operators in metrics $ H_{\mu\nu} $ and $ h_{\mu\nu} $, and making use of the second equation in (\ref{emtproj}).

We recall that in the quotient space $\cal{M} $ there exist only the spatial components of  fields and therefore one can introduce a vielbein field $ e^{\,a}_i $, such that
\begin{eqnarray}
H_{ij}& = & \delta_{ab} e^{\,a}_i e^{\,b}_j\,,~~~~\chi_i= e^{\,a}_i  \chi_a\,,
\label{eucviel}
\end{eqnarray}
where $ \delta_{ab} $ is a flat $ (\textrm{D}-1) $-dimensional Euclidean metric. Choosing, for the sake of certainty, the spinors as
\begin{eqnarray}
\chi_{a}&=& \left(
\begin{array}{c}
0 \\
\theta_a \\
\end{array}
\right)\,,~~~~\Upsilon=\left(
\begin{array}{c}
0 \\
\Upsilon_0 \\
\end{array}
\right)\,,
\end{eqnarray}
where $ \theta_a $ is a Grassmann variable, it is easy to show that the action (\ref{primeaction}) reduces to the form
\begin{eqnarray}
I=\int d\sigma \left(\frac{1}{2}\, H_{ij} \dot{x}^i \dot{x}^j + \frac{i}{2}\,\delta_{ab} \theta^a
\frac{D\theta^b}{D\sigma}\right)\,,
\label{finalaction}
\end{eqnarray}
where the overdot means the derivative $ d/d\sigma $ and  to simplify the notation we have omitted  primes, implying that all operations are taken  with respect to the effective metric (\ref{eff}). This action looks precisely the same as  that describing the dynamics of a pseudo-classical spinning point particle in a curved background \cite{gibbons}. However,  in our case  the background  is a curved Euclidean space where the length parameter $ \sigma $ plays the role
of ``time"-evolution. The corresponding constraint equations are given by
\begin{eqnarray}
 {\cal H} &=&\frac{1}{2}H_{ij} \dot{x}^i  \dot{x}^j = 1\,,~~~Q =  \dot{x}^i e_i^a \theta_a = -\Upsilon_0.
 \label{finalconstr}
\end{eqnarray}
Thus, we conclude that {\it the dynamics of a stationary spinning string in a $\textrm{D}$-dimensional stationary spacetime becomes equivalent to that of a pseudo-classical spinning point particle in a $(\textrm{D}-1)$-dimensional effective space  whose metric is obtained by conformal adjusting the metric of the quotient space, as given in equation }(\ref{eff}).

\section{Symmetries and Conserved Quantities}

The general theory of spacetime symmetries and conserved quantities in  supersymmetric mechanics of pseudo-classical spinning point particles  has been developed in \cite{gibbons,holten}.  Our result in the previous section shows that this theory can be used to describe the symmetries of stationary spacetimes in terms of the motion of  stationary spinning strings by adopting it to the effective (one dimension fewer) Euclidean space. Since in the latter case the metric
involves a conformal factor, as in (\ref{eff}),   not all {\it nongeneric} (hidden symmetries)  of the original spacetime  may survive in the effective space. That is, the explicit form of the  conformal factor plays a crucial role for the existence of nongeneric symmetries in the motion of the stationary spinning string \cite{aa1}. Following \cite{gibbons}, we  give here the basic equations describing the symmetries and  conserved quantities   which are admitted  by the action (\ref{finalaction}).  This action gives rise to the covariant  momentum
\begin{eqnarray}
 \Pi_i &=& p_i + \frac{i}{2}\, \omega_{i a b } \theta^a \theta^b= H_{ij} \dot{x}^j\,,
\label{covmom}
\end{eqnarray}
which enables one to pass to the  Hamiltonian description of the theory in terms of covariant phase-space variables $ (x^i\,, \Pi_i\,, \theta^a) $.  The associated Hamiltonian has the form
\begin{equation}
{\cal H}=\frac{1}{2}\,H^{ij}\,\Pi_i \Pi_j\,,
\label{ham}
\end{equation}
and one can define the evolution of any phase-space function $ {\cal J}( x, \Pi, \theta) $ in terms of its  Poisson-Dirac bracket with this Hamiltonian. We have
\begin{equation}
\frac{d {\cal J}}{d\sigma}= \{{\cal J}, {\cal H}\}\,.
\label{evol1}
\end{equation}
The Poisson-Dirac bracket of two general phase-space  functions is given by
\begin{equation}
 \{ F, G\}={\cal D}_i\, F \frac{\partial G}{\partial \Pi_i}-\frac{\partial F}{\partial \Pi_i}\,{\cal D}_i G
 -R_{ij} \,\frac{\partial F}{\partial \Pi_i}\frac{\partial B}{\partial
 \Pi_j}+i(-1)^{\epsilon (F)}\,\frac{\partial F}{\partial \theta^a}\frac{\partial G}{\partial
\theta_a}\,\,,
 \label{pdb}
\end{equation}
where $ {\epsilon (F)} $ stands for the Grassmann parity of  $ F $, the phase-space covariant derivative operator is defined as
\begin{equation}
{\cal D}_i F = \partial_i F +\Gamma_{ij}^k \,\Pi_k \,\frac{\partial F}{\partial \Pi_j}+\omega^{\ a}_{i\ \ b} \,\theta^b
\frac{\partial F}{\partial\theta^a}\,\,,
\label{pscovd}
\end{equation}
and the spin-valued curvature tensor
\begin{equation}
R_{ij}=\frac{i}{2}\,\theta^a \theta^b R_{a b\, ij}\,.
\label{rie}
\end{equation}
Let us suppose that the set of phase-space functions $ {\cal J}( x, \Pi, \theta) $ describes all symmetries or associated conserved quantities of the system. Then from equation (\ref{evol1}), it follows that they all must commute with the Hamiltonian (\ref{ham}). That is, we have the equation
\begin{equation}
 \{{\cal J}, {\cal H}\}=0\,,
\label{vanishpd}
\end{equation}
Next, we expand the conserved quantities of the motion in terms of the covariant momentum as follows
\begin{equation}
{\cal J}=\sum_{n=0}^\infty \frac{1}{n!} \,J^{(n)}_{ i_1 \dots  i_n}(x,\, \theta)\,  \Pi^{i_1} \dots \Pi^{i_n}\,,
\label{consts}
\end{equation}
where the components $ J^{(n)}_{ i_1 \dots  i_n}(x,\, \theta) $ can be thought of as Killing tensors generating the symmetries of the action (\ref{finalaction}). Substituting  this  expansion in equation (\ref{vanishpd}), we obtain the chain of equations
\begin{equation}
D_{(i_{n+1}} J^{(n)}_{i_1 \dots i_n)} + \frac{\partial J^{(n)}_{(i_1 \dots i_n}}{\partial\theta^a}\,\omega^{\ a}_{i_{n+1)}
   b }\,\theta^b=
    R_{j (i_{n+1}}
    J^{(n+1)j}_{i_1 \dots i_n)}\,,
    \label{chain}
\end{equation}
which couple the Killing tensors of different rank. In this  sense, they are of  generalized Killing equations \cite{holten}. These equations admit the most simple general solutions given by the second rank Killing tensor  $ K_{ij}= H_{ij} $ and the Grassmann-odd Killing vector $ I^a=\theta^a $. In the first case we have the Hamiltonian (\ref{ham}), describing the translational symmetries along the evolution parameter $ \sigma $, while the later case gives rise to the supercharge  $ Q= \Pi_a \theta^a $ (see also (\ref{finalconstr})), generating supersymmetry transformations. Along with their corresponding duals, these symmetries  form  the set of all {\it generic} symmetries which are actually  built in the  action (\ref{finalaction}). In particular, the supercharge  $ Q $ generates
the supersymmetry transformations
\begin{eqnarray}
\delta x^i & =& i \epsilon \{Q, x^i\}=- i \epsilon e^i_{\,a} \theta^a\,, ~~~~~~
\delta \theta^a = i \epsilon \{Q, \theta^a\}=
\epsilon e^{\,a}_i  \dot{x}^i+\delta
  x^i \omega^{\ a}_{i\ \ b}\,\theta^b\,,
\label{supertrans}
\end{eqnarray}
where $ \epsilon $ is the infinitesimal, Grassmann-odd parameter of the transformations. The corresponding superalgebra is given by
\begin{eqnarray}
\{ Q, {\cal H}\}& = & 0\,, \ \ \  \{Q, Q\}=-2 i {\cal H}\,.
\label{superalg}
\end{eqnarray}
The existence of nongeneric symmetries  depends, in general, on the   form of the metric (\ref{eff}) and in each particular case one needs to solve  equations (\ref{chain}) explicitly.  If one supposes that the theory also admits  a nongeneric supersymmetry transformation
\begin{eqnarray}
\delta x^i & = &- i \epsilon f^i_{\,a} \theta^a \equiv -i \epsilon J^{(1) i}\,,
\label{nongsuper}
\end{eqnarray}
it is then straightforward to show that this transformation is generated by the supercharge
\begin{equation}
Q_{f}=f^i_{\,a}\Pi_i \theta^a +\frac{i}{6}\,C_{abc} \theta^a \theta^b \theta^c\,,
\label{ngenscharge}
\end{equation}
where the tensors $ f^i_{\,\,a} $ and  $ C_{abc} $ obey the conditions
\begin{eqnarray}
D_i f_{j a}+D_{j} f_{i a} & = &0\,,~~~~D_i C_{abc}=-\left(R_{i j b c}f^j_{\,\,a}+ R_{i j c a}f^j_{\,\,b}+ R_{ij a b}f^j_{\,\,c}\right)
\label{feq}
\end{eqnarray}
which are obtained from equations (\ref{chain}). It is important to note that, unlike the case of two original supercharges $ Q $ in (\ref{superalg}),  the Poisson-Dirac bracket of two supercharges $ Q_{f} $ does not close on the Hamiltonian. Instead, it defines  the corresponding Killing tensor, Killing vector and Killing scalar, thereby forming an {\it exotic} algebra \cite{gibbons}. On the other hand, if we require the tensor  $ f_{ij}=f_{i a} e_j^{\,\,a} $ be antisymmetric, then  the Poisson-Dirac bracket of $ Q_{f} $  with $ Q $ vanishes. In this case, the defining equation for the third rank tensor $ C_{abc} $ has the most simple form \cite{gibbons}
\begin{equation}
C_{abc}= - 2 e^i_{\,a}  e^j_{\,b} e^k_{\,c}\, D_i f_{jk}\,,
\label{cdef}
\end{equation}
and the superinvariant $ Q_{f} $  is determined by the second rank Killing-Yano tensor $ f_{ij} $.

\section{Stationary Spinning Strings in the Kerr-Newman spacetime}

In this section, as an illustrative example, we consider the symmetries and conserved quantities in terms of the motion of a stationary spinning  string near a black hole which is given by the Kerr-Newman spacetime metric
\begin{eqnarray}
ds^2 & = & -{\Delta\over \rho^2}\,(dt - a \sin^2\theta\,d\phi)^2 + \rho^2 \left(\frac{dr^2}{\Delta} +
d\theta^{2}\right)
+\,{\sin^2\theta\over \rho^2}\,
\left[a dt - (r^2+a^2) d\phi \right]^2 ,
\label{kn}
\end{eqnarray}
where the metric functions
\begin{eqnarray}
\Delta &=& r^2+a^2-2Mr+Q^2\,,~~~~~~ \rho^2 = r^2+a^2\cos^2\theta
\end{eqnarray}
and the parameters $ M $, $ a=J/M $ and $ Q $ are the mass, rotation parameter and  the electric charge of the black hole, respectively.

According to the theory developed above, the motion of the stationary spinning string becomes equivalent to that of a spinning point particle in the effective three-dimensional space with the metric  (\ref{eff}). That is, we have the space interval
\begin{equation}
dl^2=
\frac{\Sigma}{\Delta}\,dr^2+
    \Sigma\, d\theta^2+\Delta\sin^2\theta  d\phi^2\,,
    \label{3eff}
\end{equation}
where
\begin{equation}
\Sigma=\rho^2-2Mr + Q^2\,.
\end{equation}
For this metric we can consider a conserved quantity of the form
\begin{equation}
{\cal J}=\frac{1}{2}\, K_{ij} \Pi^i \Pi^j + I_i \Pi^i.
\end{equation}
Then from equation (\ref{vanishpd}) we obtain that  $ K_{ij} $ is a  Killing tensor and  $ I_i= (i/2) I_{i a b} \theta^a  \theta^a\, $ is a spin-valued Killing vector, which are determined by the equations
\begin{eqnarray}
D_{(k}K_{ij)}&=& 0 \,,~~~~~~ D_{(i} I_{j )a b} = R_{a b k (i }
    K_{j)}^{ \  k}\,.
    \label{ktkveqs}
\end{eqnarray}
It is straightforward to check that the Killing tensor equation admits the solution
\begin{equation}
K_{ij}dx^i dx^j = \Sigma \left(\frac{a^2 \sin^2\theta}{\Delta}\, dr^2 + \Delta
d\theta^2 \right) + \Delta \left(
\Delta+a^2\sin^2\theta \right) \sin^2\theta d\phi^2 \,.
\label{killtensor}
\end{equation}
This expression agrees with that given in \cite{fszh}. Next, it is convenient to pass to the local frame, which is given by the basis one-forms $ e^a= e^{\,a}_i  dx^i $ of the metric (\ref{3eff}), and put the second equation in (\ref{ktkveqs}) in the form
\begin{eqnarray}
D_{(a} F_{b ) c} = Z_{abc}\,,
\label{x1}
\end{eqnarray}
where
\begin{eqnarray}
F_{bc}&=& \frac{1}{2}\,\varepsilon_{c}^{ \  e f } I_{b  e f}\,, \ \ \ \ \ Z_{abc}=\frac{1}{2}\,\varepsilon_{c}^{  \ e f}R_{ e f d (a }
    K_{b)}^{ \ d}\,
\end{eqnarray}
and $ \varepsilon_{a b c}$ is the usual Levi-Civita symbol.  The nonvanishing  components of the tensor $ Z_{abc} $  are given by
\begin{eqnarray}
Z_{312} & = & \frac{M^2-Q^2}{\Sigma^2} \Delta\,,~~~~Z_{231}= \frac{M^2-Q^2}{\Sigma^2}a^2\sin^2\theta\,,~~~~Z_{123}= -( Z_{312}+Z_{231})\,,
\label{zcomps}
\end{eqnarray}
Substituting these expressions in equation (\ref{x1}) it is straightforward to show that for the tensor
\begin{eqnarray}
F_{ab}=S_{ab}+ A_{ab}\,,
\end{eqnarray}
the symmetric part $ S_{ab} $ vanishes identically, while solving the remaining equations we find that the  nonvanishing components  of  the  antisymmetric tensor $ A_{ab} $ are given by
\begin{eqnarray}
A_{13}&=& \frac{a^2\sin2\theta}{\sqrt{\Sigma}}\,\,,~~~~~ A_{23}= - 2\left (r-M\right)\sqrt{\frac{\Delta}{\Sigma}} \,\,.
\label{antisymcomps}
\end{eqnarray}
Finally, using this expressions we obtain that the spin-valued Killing vector is given by
\begin{equation}
I_i dx^i =   \frac{i  a^2 \sin 2\theta}{\sqrt{\Delta}} \left(\theta^1 \theta^2  dr - \frac{\Delta \sin\theta}{\sqrt{\Sigma}}\, \theta^2 \theta^3 d\phi \right) - 2 i (r-M) \left(\sqrt{\Delta}\, \theta^1 \theta^2 \, d\theta - \frac{\Delta \sin\theta}{\sqrt{\Sigma}}\, \theta^3 \theta^1 d\phi \right)\,.
\label{spinkillvec}
\end{equation}

It is now natural to ask whether the space (\ref{3eff}) admits a nongeneric supercharge of the form (\ref{ngenscharge}). To answer this question, one needs to look for the  Killing-Yano tensor $ f_{ij} $ of this space.  Writing down explicitly  the Killing-Yano equation
\begin{equation}
D_{(i} f_{j)k}  =0\,
\label{projyanoeq}
\end{equation}
in the background (\ref{3eff}), we see that $ f_{12}=0 $. Meanwhile, from the integrability conditions of  the equations for the remaining components $ f_{13} $ and  $ f_{23} $, we find that
\begin{eqnarray}
f_{13}& = & 0\,,~~~~~ a^2 f_{23} =0 \,.
\end{eqnarray}
It follows that a nontrivial solution exists only for the vanishing rotation parameter, $ a=0 $. This solution has the simple form
\begin{equation}
\frac{1}{2}\,f_{ij}dx^i\wedge dx^j = \left(\Delta\right)^{3/2}\sin\theta d\theta \wedge
d\phi \,.
\label{projyano}
\end{equation}
We note that in three dimensions the Killing-Yano tensor can always be expressed as  the dual of a one-form field which must be closed. In our case,  it is the dual of the conformal Killing vector field  $ \Omega= \sqrt{\Delta}\, dr $ of the space (\ref{3eff}). Evaluating now the nonvanishing components  of the third rank tensor in (\ref{cdef}), we obtain that
\begin{equation}
C_{123}=\frac{2(r-M)}{\sqrt{\Delta}}\,.
\label{c123}
\end{equation}
With the expressions (\ref{projyano}) and (\ref{c123}) the associated  supercharge in (\ref{ngenscharge}) does not generate a new nongeneric supersymmetry. It is related to the  conserved sum of the orbital and spin angular momenta of the spinning point particle. The Poisson-Dirac bracket of two such supercharges gives the  Killing tensor (reducible)  and the spin-valued Killing vector which  agree precisely with  the $ a=0 $ limit of the expressions in (\ref{killtensor}) and (\ref{spinkillvec}).

\section{Conclusion}

The main purpose of this paper was to show that in a stationary spacetime, the dynamics of  stationary spinning strings governed by
a worldsheet supersymmetric action becomes equivalent to that of
pseudo-classical spinning point particles in the  effective metric of the quotient space of the original spacetime.  Employing the Geroch procedure of foliation of the spacetime by its Killing trajectories, we have shown that the action of  a  stationary spinning string  reduces to the action of a  spinning point particle in the quotient space with conformally scaled metric. This fact generalizes the similar result obtained earlier \cite{fszh} for  the stationary bosonic string dynamics and makes it possible to use the known general theory of spacetime symmetries in mechanics of  spinning point particles \cite{gibbons}. In this framework, we have explored the symmetries  of the Kerr-Newman spacetime in terms of the motion of the stationary spinning string. We  have solved the equations for  symmetries in the three-dimensional effective quotient space of the Kerr-Newman spacetime and presented the explicit expressions for the Killing tensor as well as for the spin-valued Killing vector. We  have also shown that the effective space does not admit the Killing-Yano tensor, except in the case of vanishing rotation parameter. In the latter case, the associated supercharge, unlike the case of  the original spacetime, does not generate a new nontrivial supersymmetry. It becomes  related to the conserved sum of the orbital and spin angular momenta of the spinning point particle.

\section{Acknowledgments}
The authors thank  Nihat Berker and Teoman Turgut for their  encouragements and support.


\begin{thebibliography}{99}

\bibitem{fszh}  V. P. Frolov, V. D. Skarzhinsky, A. I. Zelnikov and O. Heinrich,  Phys. Lett. B {\bf 224} (1989) 255.
\bibitem{carter}  B. Carter,  Phys. Rev.  D {\bf 174} (1968)  1559.
\bibitem{wp} M. Walker and R. Penrose, Commun. Math. Phys. {\bf 18} (1970) 265.
\bibitem{fk1}  V. P. Frolov and D.  Kubiz\v{n}\`{a}k,  Phys. Rev. Lett. {\bf 98}  (2007) 011101.
\bibitem{fk2}  D.  Kubiz\v{n}\`{a}k and V. P. Frolov,  Class. Quant. Grav. {\bf 24} (2007)  F1.
\bibitem{fk3}  V. P. Frolov and D.  Kubiz\v{n}\`{a}k ,  Class. Quant. Grav. {\bf 25} (2008) 154005.
\bibitem{fk4}  D.  Kubiz\v{n}\`{a}k and V. P. Frolov, J. High Energy Phys. {\bf 0802} (2008) 007.
    \bibitem{cclp} Z-W. Chong , M. Cvetic, H. L\"u  and C. N.
Pope,  Phys. Rev. Lett. {\bf 95} (2005) 161301.
\bibitem{dkl}    P. Davis, H. K. Kunduri and J. Lucietti, Phys. Lett. B {\bf 628} (2005) 275.
\bibitem{ad} A. N. Aliev and O. Delice, Phys. Rev. D {\bf  79} (2009) 024013.
\bibitem{aa1} H. Ahmedov and A. N. Aliev, Phys. Rev. D {\bf  78} (2008) 064023.
\bibitem{chandra} S. Chandrasekhar, Proc. R. Soc. London A {\bf 349} (1976) 571.
\bibitem{guven}  R. G\"{u}ven, Phys. Rev. D {\bf  16} (1977) 1706.
\bibitem{penrose} R. Penrose, Ann.  N.Y. Acad. Sci. {\bf 224} (1973) 125; R. Floyd, Ph.D. thesis, London Univ., 1973.
\bibitem{cartermc}  B. Carter and R. G. McLenaghan, Phys. Rev. D {\bf  19} (1979) 1093.
\bibitem{gibbons} G. W. Gibbons, R. H. Rietdijk and J. W. van Holten, Nuclear Phys. B {\bf 404} (1993) 42.
\bibitem{aa2} H. Ahmedov and A. N. Aliev, in preparion  (2009).
\bibitem{brink} L. Brink, P. Di Vecchia and P. Howe, Phys. Lett. B {\bf 65} (1976) 471.
\bibitem{deser} S. Deser and B. Zumino,  Phys. Lett. B {\bf 65} (1976) 369.
\bibitem{green} M. B. Green, J. H. Schwarz and E. Witten, {\it
Superstring Theory v.1}  (Cambridge Univ. Press, Cambridge, England, 1997).
\bibitem{geroch} R. Geroch,  J. Math. Phys. {\bf 12} (1971) 918.
\bibitem{holten}  R. H. Rietdijk and J. W. van Holten, Class. Quant. Grav. {\bf 7} (1990) 247; R. H. Rietdijk, Ph.D. thesis, Amsterdam Univ., 1992.

\end{thebibliography}
\end{document}